\documentclass[twocolumn,floatfix,showpacs,preprintnumbers,amsmath,amssymb,aps,prl]{revtex4}
\usepackage{amssymb,amsmath,graphics,epsfig}
\begin{document}

\title{Radiative force from optical cycling on a diatomic molecule  }
\author{E.S. Shuman, J.F. Barry, D.R. Glenn, and D. DeMille}
\affiliation{Department of Physics, Yale University, PO Box 208120, New Haven, CT 06520, USA}
\date{\today}

\begin{abstract}
We demonstrate a scheme for optical cycling in the polar, diatomic molecule strontium monofluoride (SrF) using the $X ^2\Sigma^+\!\rightarrow\!A^2\Pi_{1/2}$ electronic transition.  SrF's highly diagonal Franck-Condon factors suppress vibrational branching.  We eliminate rotational branching by employing a quasi-cycling $N\!=\!1\!\rightarrow\! N^\prime\!=\!0$ type transition in conjunction with magnetic field remixing of dark Zeeman sublevels.   We observe cycling fluorescence and deflection through radiative force of an SrF molecular beam using this scheme.  With straightforward improvements our scheme promises to allow more than $10^5$ photon scatters, possibly enabling the direct laser cooling of SrF.

\end{abstract}

 \maketitle

 \clearpage

Recently there has been substantial interest in ground state ultracold ($\lesssim$ 1 mK) polar molecules, which have possible applications to a wide range of topics \cite{Carr09}.  Samples of ultracold molecules could prove tremendously valuable for tests of fundamental symmetries \cite{Tarbutt09,Flambaum07,DeMille08b}, for simulation of solid state systems \cite{Goral02,Micheli06,Barnett06}, as qubits in a quantum computer \cite{DeMille02}, and for studying ultracold chemistry\cite{Balakrishnan01,Krems08}.  Currently the best technique for creating ultracold molecules relies on their assembly from preexisting ultracold atoms \cite{Sage05,Ni08}, and recent experiments have produced ground state polar molecules near quantum degeneracy \cite{Ni08}.  Unfortunately only bialkali molecules have been produced this way and the number of molecules created is fairly small ($\sim 10^4$) \cite{Ni08}.  There is substantial interest in developing techniques for direct cooling of molecules, enabling use of species with different structures (e.g. unpaired electron spins, which are important for several proposed applications \cite{Tarbutt09,Andre06,Micheli06,DeMille08b}), and possibly also much larger samples.   Molecules have been directly cooled using a cryogenic buffer gas \cite{Weinstein98} or via supersonic expansion followed by beam deceleration \cite{Bethlem00}, but so far the temperature reached with these techniques is limited to $\sim\! 10$ mK.

The prospect of direct laser cooling of molecules is attractive because large numbers could in principle be obtained for new types of species at ultracold temperatures.  Unfortunately, molecules possess rotational and vibrational degrees of freedom resulting in decays to unwanted sublevels, so finding the closed cycling transitions required for laser cooling is a challenge.  However, recent proposals have shown that certain molecules may be amenable to laser cooling because their Franck-Condon factors (FCFs) suppress decays to excited vibrational states \cite{DiRosa2004,Stuhl08}.  Furthermore Stuhl \textit{et al.} pointed out that a $F\!=\!1\rightarrow F^\prime\!=\!0$ type transition eliminates decays to all but the initial $F\!=\!1$ state \cite{Stuhl08}.  (We use $N$, $v$, $I$, $S$, $F$, $M$, and $P$ as the rotational, vibrational, nuclear spin, electronic spin, total angular momentum, Zeeman, and parity quantum numbers respectively.)  In this scheme dark Zeeman levels in the ground state must be remixed to obtain optical cycling \cite{Stuhl08,Berkeland2002}.

\begin{figure}
\includegraphics[height=2.2in]
{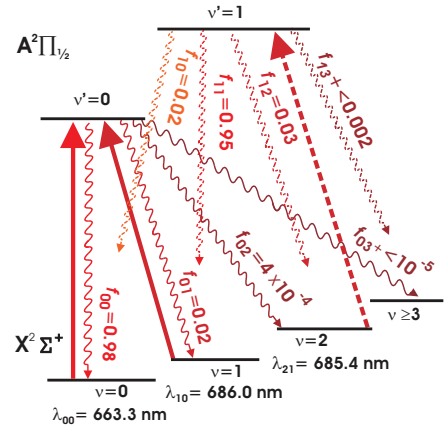} \caption{Relevant electronic and vibrational structure in SrF (color online).  Solid upward lines indicate laser-driven transitions in the experiment at the wavelengths $\lambda_{v,v'}$.  Solid wavy lines indicate spontaneous decays from the A($v\!=\!0$) state with FCFs $f_{0v}$ as shown.  Dashed wavy lines indicate spontaneous decays from the A($v\!=\!1$) state populated by the proposed second vibrational repump laser (dashed vertical arrow), with FCFs $f_{1v}$ as shown.} \label{fig1}
\end{figure}

Here we propose and demonstrate a scheme for optical cycling that uses the X$^2\Sigma^+_{1/2}\rightarrow$ A$^2\Pi_{1/2}$ electronic transition in the molecule SrF.  This transition has the required highly diagonal FCFs.  Use of a P branch $N\!=\!1\!\rightarrow\! N^\prime\!=\!0$ type rotational transition eliminates rotational branching, while a magnetic field remixes the dark Zeeman sublevels.  The X$(N\!=\!1)$ state of SrF has four hyperfine structure (HFS) levels, so we outline a technique for simultaneously addressing all of these levels using a single laser.  We apply this scheme to a cryogenic helium-cooled molecular beam of SrF, and observe both cycling fluorescence and deflection of the beam by radiative force.  With only one main pump laser and one vibrational repump laser, our observations correspond to the scattering of $\sim\!150$ photons from each SrF molecule.  These results suggest that addition of a second vibrational repump laser should allow more than $10^5$ photon scattering events, given sufficient interaction time.  This would be sufficient to enable the direct laser cooling of SrF.

The electronic, vibrational, and rotational structure of SrF \cite{Herzberg79} make it attractive for potential laser cooling.  The first electronically excited state A$^2\Pi_{1/2}$ has lifetime $\tau\!=\!24$ ns (natural width $\Gamma_{n}\!=\!2\pi\!\times\!7$ MHz) \cite{Dagdigian1974} and can only radiatively decay back to the ground  X$^2\Sigma^+_{1/2}$ state \cite{Allouche93}.  SrF's ground state dipole moment of $\mu\!=\!3.5$ D \cite{Ernst1985} and unpaired electron spin also make it useful for several of the previously mentioned applications.  We have calculated FCFs for the X$^2\Sigma^+_{1/2}\rightarrow$ A$^2\Pi_{1/2}$  transition using a Morse potential derived from molecular constants \cite{Herzberg79,Steimle93,Steimle2008}.  As shown in Fig. 1, the branching to $v\geq3$ states is $f_{03^+}\!<\! 10^{-5}$, so using the X($v\!=\!0$)$\rightarrow$ A($v^\prime\!=\!0$) transition as the main pump, the X($v\!=\!$1)$\rightarrow$ A($v^\prime\!=\!0$) transition as the first vibrational repump, and the X($v\!=\!2$)$\rightarrow$ A($v^\prime\!=\!1$) transition for the second vibrational repump should result in $N_{scat}\!=\!1/f_{03^+}\!>\!10^5$ photons scattered before higher vibrational levels are populated.  Conveniently, these transitions occur at frequencies where ample power is available in laser diodes.

By driving  $N\!=\!1\rightarrow N^\prime\!=\!0$ transitions, only decays back to the $N^{\prime\prime}\!=\!1$ level are allowed.  Unlike in the X$^2\Sigma^+_{1/2}$ state (described by Hund's case b), in the excited A$^2\Pi_{1/2}$ state (Hund's case a), $N$ is not generally a good quantum number.   Here states are instead labeled with the coupled rotational plus electronic angular momentum, $J$ \cite{Brown03}.  Nevertheless in the $J\!=\!1/2^+$ state, projection onto the Hund's case b basis shows that $N$ is well-defined.  As shown in Fig. 2a, the ground state of SrF has both spin-rotation ($S\!=\!1/2$) and HFS ($I\!=\!1/2$) interactions which split the $N=1$ level into four components \cite{Childs81}.  HFS is unresolved in the A$^2\Pi_{1/2}$ states \cite{Kandler89}.  The HFS interaction is not diagonal in either $N$ or $J$, so dipole selection rules require only that $\Delta F\!=\!0,\pm1$ and $\Delta P\!=\!\pm1$.  Transitions from any individual $N\!=\!1$ HFS level then result in population distributed over the other three HFS levels, so all four HFS levels must be pumped simultaneously.  The parity selection rule prevents decay from A$( J\!=\!1/2^+)$ state to any of the X$(N\!=\!0,2)$ levels.  HFS mixing between the X($N\!=\!3, F\!=\!2$) and ($N\!=\!1, F\!=\!2$) states, or between the A($J\!=\!1/2, F\!=\!1$) and ($J\!=\!3/2, F\!=\!1$) states allows decays to the X ($N\!=\!3$) states with small branching ratio $\eta$.  $\eta$ can be estimated using second-order perturbation theory, and in both cases $\eta \lesssim 10^{-6}$.  Decays to all other rotational levels are forbidden due to the $\Delta F$ selection rules.

\begin{figure}
\includegraphics[height=2.3in]
{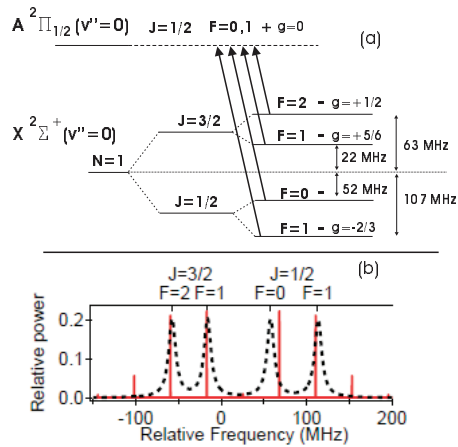} \caption{(a) Relevant rotational energy levels, splittings, parity, and magnetic $g$-factors in the SrF cycling scheme.  The relevant cycling transitions are shown as vertical arrows.  (b)  Laser (vertical lines) and molecular (dashed curve) spectra for addressing all HFS sublevels.  Molecular lines are shown with a power broadened linewidth of 10 MHz and are equally weighted for clarity.  For an EOM modulation index of $m\!=\!2.6$ and frequency of 42.5 MHz, there is a laser sideband $<10$ MHz detuned from each molecular line.} \label{fig2}
\end{figure}

Conveniently, each X($N\!=\!1$) HFS level of SrF can be pumped using a single laser and an electro-optic modulator (EOM).  As shown in Fig. 2b, a laser of frequency $f_o$, passing through an EOM phase modulated at 42.5 MHz with a modulation index of $m\!=\!2.6$, produces a laser spectrum that nearly matches all four X$(N\!=\!1)\rightarrow$ A$(J\!=\!1/2)$ transitions.  For a laser tuned as shown in Fig. 2 the detuning from each HFS level is $<10$ MHz.

A significant problem is the presence of dark Zeeman sublevels. For instance, if a $z$-polarized laser is used to drive the X$\rightarrow$A transitions, then the X$(N\!=\!1,F\!=\!2,M\!=\!\pm2)$ sublevels are dark.  Due to HFS interactions, we calculate that only 3 photons can be scattered on average before molecules are pumped into dark states, terminating the cycling.  Because the X state has a magnetic moment of $\mu\sim\! \mu_B\!=\!1.4$ MHz/Gauss, a magnetic field of a few Gauss at an angle $0\!<\!\theta_B\!<\! 90^\circ$ relative to the laser polarization can remix these levels back into the cycle \cite{Berkeland2002,Kloter2008}.   The remixing rate is limited to a few MHz to avoid broadening the molecular transitions and hence reducing the photon scattering rate.  With sufficient Zeeman mixing our scheme will only be limited by decays to higher vibrational levels, so scattering more than $10^5$ photons should be possible.

We implement our cycling scheme using a cryogenic buffer-gas cooled molecular beam of SrF \cite{Maxwell05}.  A copper cell ($2.5\!\times\!2.5\!\times\!2.5$ cm) is attached to the 4K surface of a liquid helium-cooled cryostat held under vacuum at $\sim\! 1\times10^{-6}$ torr.  4 K He buffer gas flows continuously into the cell, leading to a steady state density of $\sim\!10^{15}$ cm$^{-3}$.  To create SrF we ablate a solid, pressed, SrF$_2$ target located inside the cell using 5 ns, $\sim$20 mJ pulses of 1064 nm light from a Q-switched Nd:YAG laser typically operating at 1 Hz.  After sufficient collisions with the buffer gas, the SrF molecules thermalize to $\sim\!4$ K.  To form a beam, an exit aperture of radius 3 mm is placed on the face of the cell through which a mixture of He and SrF escape.  The density of He is large enough that the both the extraction efficiency and forward velocity are hydrodynamically enhanced \cite{Maxwell05,Patterson07}.  By measuring Doppler profiles we find the forward velocity of the beam to be $v_{||}\!=\!200\!\pm\!30$ m/s with a spread of $\Delta v_{||}\!\approx\!40$ m/s, and the transverse velocity spread $\Delta v_\perp\approx30$ m/s.  The beam has $\sim\!10^9$ molecules in the X($v\!=\!0$, $N\!=\!1$) state per ablation shot.

\begin{figure}
\includegraphics[height=1.5 in]
{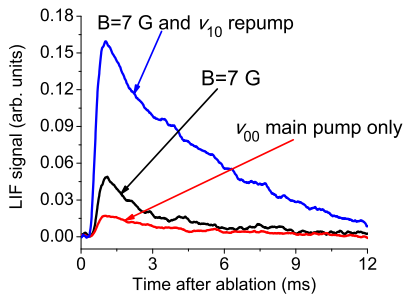} \caption{Signals demonstrating cycling fluorescence in SrF.  LIF signal from the $v_{00}$ pump laser in the pump region with $B\!=\!0$ G, with $B\!=\!7$ G, and with both the $v_{10}$ repump and $B\!=\!7$ G.  The addition of the magnetic field $B$ results in a $\sim\!3.5\times$ enhancement in signal due to remixing from dark Zemman sublevels.  The addition of the $v_{10}$ repump results in another $\sim\!3.5\times$ enhancement, indicating pumping to and from the X ($v\!=\!1$) vibrational level as a result of optical cycling.} \label{fig3}
\end{figure}

The beam passes through a 6 mm diameter hole in a coconut charcoal-covered 4K copper plate, 2.5 cm from the cell. This plate acts as an absorptive skimmer limiting He flow into the rest of the apparatus \cite{Tobin1987}.  The beam then exits the cryostat and passes through a 1.5$\times$3 mm slit which collimates the beam transversely to $\Delta v_\perp\simeq\! 3$ m/s.  The room temperature experimental apparatus consists of a pump and a probe region, separated by $D\!=\!12.5$ cm.  The pump region is defined by two pairs of orthogonal 4 cm diameter Helmholtz coils used to add a magnetic field $\mathbf{B}$, and a 3 cm clear aperture which gives optical access to the SrF beam.  The main $v_{00}$ pump and $v_{10}$ repump lasers have linear polarization, 42.5 MHz sidebands shown in Fig. 2b, diameters of 1.1 mm (FWHM), and powers of 50 and 60 mW respectively.  They are spatially overlapped before passing through the pump region along the 3 mm axis of the slit.  The laser beams are reflected around a circular path for a total of 8 passes through the molecular beam, all originating from the same direction. This results in in $l_0\!=\!0.9$ cm and $t_0=v_{||}/l_0\!=\!44$ $\mu$s of interaction length and time.  Laser-induced fluorescence (LIF) from the $v_{00}$ transition is monitored in the pump region using a photomultiplier tube.  The position of the SrF beam in the probe region is determined by imaging LIF from the X$(N\!=\!1)\rightarrow$ A$(J\!=\!1/2)$ transition onto an intensified CCD camera. A  ``cleanup'' $v_{10}$ laser is introduced between pump and probe regions, to return any residual population in the X$(v\!=\!1)$ state back to the X($v\!=\!0$) state.  Both probe and cleanup lasers have 42.5 MHz sidebands to maximize the LIF, are retroreflected to eliminate artificial Doppler shifts, and are linearly polarized.

\begin{figure}
\includegraphics[height=2.0in]
{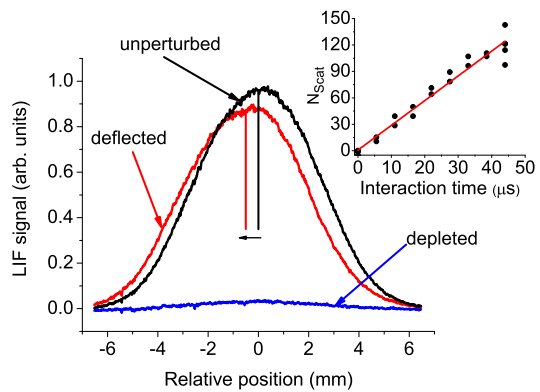} \caption{Radiative force deflection of the SrF beam.  Positional LIF from the $v_{00}$ probe laser with no lasers in the pump region (unperturbed), with $v_{00}$ pump and $v_{10}$ repump lasers in pump region (deflected), and with pump laser but no repump or $v_{10}$ cleanup laser  (depleted).  Without the repump and cleanup lasers, nearly complete depletion of the $v\!=\!0$ sublevel is observed.   Addition of these lasers results in nearly full recovery of the LIF signal amplitude, indicating optical pumping to and repumping from the $v\!=\!1$ level.  The pump and repump lasers also cause deflection of the SrF beam due to radiative force.  This plot shows raw data, with no rescaling applied.  Inset: Dependence of the number of scattered photons on the interaction time in the pump region.  Points are data and the line is a fit, showing that $N_{scat}$ increases linearly with interaction time.} \label{fig4}
\end{figure}

In Fig. 3 we show LIF from the main pump transition in the interaction region.  The addition of $B = 7$ G (oriented at angle $\theta_B \!=\! 30^\circ$ relative to the linear laser polarization) increases the LIF by a factor of 3.5.  We find that the magnitude of the LIF signal does not depend strongly on the magnitude or direction of $\mathbf{B}$, as long as $B\gtrsim$ 3G and $\theta_B\neq0,90^\circ$. Assuming complete remixing of the dark Zeeman states, each molecule should scatter $N_{00} \!\sim\! \frac{1}{f_{10}}$ photons.  In a separate experiment, we determined $f_{10}$ from absorption spectroscopy on the X($v\!=\!1, N\!=\!1$)$\rightarrow$ A ($v^\prime\!=\!1,0$) transitions in the buffer cell.  From the ratio of observed transition strengths, and using the calculated value $f_{11}\!=\!0.95$ (Fig. 1), we determined $f_{10}\!=\!0.021 \pm .005$ and correspondingly $ N_{00} \!=\! 48$, in good agreement with our calculations.   The LIF enhancement on application of $\mathbf{B}$ indicates that $\sim\! 48/3.5\!=\!14$ photons are scattered when no $B$-field is applied.  Although for $B\!=\!0$ we expect only 3 photon scatters before pumping into a dark Zeeman sublevel, the earth's $B$-field (not canceled in our measurements) is sufficient to account for our observations.

The addition of the $v_{10}$ repump laser further increases the LIF by a factor of 3.5, indicating a total of number of photons scattered: $N_{scat}\! \approx\! 170$.  This represents a significant radiative force on the molecules and, as shown in Fig. 4, results in a substantial deflection of the SrF beam.  The unperturbed SrF beam is approximately 6 mm wide in the probe region as the result of the collimating slit.  The addition of the $v_{00}$ and $v_{10}$ lasers causes a shift in position by $\sim\!0.5$ mm while the width remains $\sim\!$ 6 mm.  We also show the LIF signal in the probe region without the $v_{10}$ repump or cleanup lasers.  Here only 5\% of the molecules remain in the X($v\!=\!0$) state.  The addition of the $v_{10}$ lasers recovers 90\% of LIF signal, indicating nearly complete pumping to and repumping from the X($v\!=\!1$) state.  The remaining 10\% loss is in reasonable agreement with the expected loss of $\sim\! 6$\% of the population to the X$(v\!=\!2)$ level (not repumped in this experiment) after 170 photons are scattered.

We independently determine $N_{scat}$ from the observed deflection.  Each scattered photon changes the transverse velocity by the photon recoil velocity $v_{r}=\hbar/M \lambda=5.6$ mm/s, where $M=107$ amu is the mass of SrF and $\lambda\!=\!663$ nm is the wavelength of the $v_{00}$ transition.  A deflection of the SrF beam $d$ then corresponds to $d\!=\!\frac{N_{scat}v_r D}{v_{||}}$.  The observed deflection $d\!=\!0.5$ mm corresponds to $N_{scat}\!=\!140$ which is in reasonable agreement with the number estimated from the LIF increase.

We also  measured the deflection as a function of $t_0$, as shown in the inset in Fig. 4.  By varying the number of passes in the pump region, we change $t_0$ from 0 to 44 $\mu$s in eight 5.5 $\mu$s increments.   These data clearly show a linear increase for $N_{scat}$ versus $t_0$.  Fitting the linear dependence yields a scattering rate $\Gamma_{scat}\!=\!3$ MHz.  We can compare this to the scattering rate expected assuming full saturation of the X($v\!=\!0[1]$)$\rightarrow$A($v\!=\!0[0]$) transitions and complete $B$-field remixing of ground state sublevels.  Here the molecule spends an equal amount of time in the 24 X ($v=0,1$) sublevels and the 4 A state sublevels.  Spontaneous emission requires one lifetime $\tau$ in the excited states, so the maximum scattering rate is $\Gamma_{max}\!=\!\frac{4}{24+4}\!\times\!1/\tau\!=\!6\!\times\!10^6$ s$^{-1}$.  We find $\Gamma_{scat}\!\simeq\!\Gamma_{max}/2$, indicating that all transitions are not saturated. We believe this may be due to the $\sim\! 10$ MHz detuning from the X $F=0$ transition.

Given the clear evidence for both optical cycling and radiative force in SrF, we believe it should be possible to apply these techniques to the laser cooling of SrF.  The combination of detuning the laser frequencies below resonance for all HFS lines, retroreflecting the $v_{00}$ and $v_{10}$ lasers, and applying them over sufficient interaction length should result in large transverse cooling forces. Furthermore, counterpropagating the lasers relative to the SrF beam with a suitable frequency chirp should bring the beam to a stop, as needed for loading into a trap, after scattering $N_{stop}\!\simeq\!\frac{v_{||}}{v_{r}}\!=\!3.6\!\times\!10^4$ photons.

Finally, we speculate about constructing a magneto-optical trap (MOT) for SrF.  A $B$-field alone cannot be at the proper angle to the 3 MOT lasers simultaneously, so another technique for remixing dark Zeeman sublevels must be used.  Stuhl \textit{et al.} proposed using weak DC electric field pulses in polar molecules with closely spaced $\Omega$-doublets \cite{Stuhl08}.  A more general technique for remixing the dark states in polar molecules (not relying on $\Omega$-doublet substructure) is to apply a microwave field, polarized perpendicular to the local MOT magnetic field, at the X $N\!=\!1\rightarrow N^\prime\!=\!0$ transition frequency.  In such a microwave-remixed MOT (MR-MOT), the bright and dark $N\!=\!1$ states mix via the $N\!=\!0$ levels.  For SrF, the rotational splitting of this transition is 15.0 GHz, and microwave field intensities of only $\lesssim\! 0.1$ W/cm$^2$ are sufficient to generate remixing frequencies of a few MHz.

In conclusion, we report on the demonstration of optical cycling resulting in radiative force in a diatomic molecule.  Our results are consistent with the scattering of $\sim\! 150$ photons, limited only by the finite interaction time with the lasers.  These results suggest a clear path to direct laser cooling of SrF or other species with similar structure.

We acknowledge the contributions of S. Falke, D. Farkas, P. Orth, D. Patterson, J. Petricka, M. Steinecker, A. Vutha, and C. Yale.  This work is supported by ARO, NSF, and AFOSR.

\bibliography{ColdMoleculesRefs_current}

\begin{thebibliography}{31}
\expandafter\ifx\csname natexlab\endcsname\relax\def\natexlab#1{#1}\fi
\expandafter\ifx\csname bibnamefont\endcsname\relax
  \def\bibnamefont#1{#1}\fi
\expandafter\ifx\csname bibfnamefont\endcsname\relax
  \def\bibfnamefont#1{#1}\fi
\expandafter\ifx\csname citenamefont\endcsname\relax
  \def\citenamefont#1{#1}\fi
\expandafter\ifx\csname url\endcsname\relax
  \def\url#1{\texttt{#1}}\fi
\expandafter\ifx\csname urlprefix\endcsname\relax\def\urlprefix{URL }\fi
\providecommand{\bibinfo}[2]{#2}
\providecommand{\eprint}[2][]{\url{#2}}

\bibitem[{\citenamefont{Carr et~al.}(2009)\citenamefont{Carr, DeMille, Krems,
  and Ye}}]{Carr09}
\bibinfo{author}{\bibfnamefont{L.}~\bibnamefont{Carr \textit{et al.}}},
   \bibinfo{journal}{New  J. Phys.} \textbf{\bibinfo{volume}{11}}, \bibinfo{pages}{055049}
  (\bibinfo{year}{2009}).

\bibitem[{\citenamefont{Tarbutt et~al.}(2009)\citenamefont{Tarbutt, Hudson,
  Sauer, and Hinds}}]{Tarbutt09}
\bibinfo{author}{\bibfnamefont{M.}~\bibnamefont{Tarbutt \textit{et al.}}},
  \bibinfo{journal}{Faraday Discuss.} \textbf{\bibinfo{volume}{142}},
  \bibinfo{pages}{37} (\bibinfo{year}{2009}).

\bibitem[{\citenamefont{Flambaum and Kozlov}(2007)}]{Flambaum07}
\bibinfo{author}{\bibfnamefont{V.~V.} \bibnamefont{Flambaum \textit{et al.}}} ,
  \bibinfo{journal}{Phys. Rev. Lett.} \textbf{\bibinfo{volume}{99}},
  \bibinfo{pages}{150801} (\bibinfo{year}{2007}).

\bibitem[{\citenamefont{DeMille et~al.}(2008)\citenamefont{DeMille, Cahn,
  Murphree, Rahmlow, and Kozlov}}]{DeMille08b}
\bibinfo{author}{\bibfnamefont{D.}~\bibnamefont{DeMille \textit{et al.}}},
  \bibinfo{journal}{Phys. Rev. Lett.}
  \textbf{\bibinfo{volume}{100}}, \bibinfo{pages}{023003}
  (\bibinfo{year}{2008}).

\bibitem[{\citenamefont{Goral et~al.}(2002)\citenamefont{Goral, Santos, and
  Lewenstein}}]{Goral02}
\bibinfo{author}{\bibfnamefont{K.}~\bibnamefont{Goral \textit{et al.}}},
  \bibinfo{journal}{Phys. Rev. Lett.} \textbf{\bibinfo{volume}{88}},
  \bibinfo{pages}{170406} (\bibinfo{year}{2002}).

\bibitem[{\citenamefont{Micheli et~al.}(2006)\citenamefont{Micheli, Brennen,
  and Zoller}}]{Micheli06}
\bibinfo{author}{\bibfnamefont{A.}~\bibnamefont{Micheli \textit{et al.} }},
    \bibinfo{journal}{Nature Phys.} \textbf{\bibinfo{volume}{2}},
  \bibinfo{pages}{341} (\bibinfo{year}{2006}).

\bibitem[{\citenamefont{Barnett et~al.}(2006)\citenamefont{Barnett, Petrov,
  Lukin, and Demler}}]{Barnett06}
\bibinfo{author}{\bibfnamefont{R.}~\bibnamefont{Barnett \textit{et al.}}},
  \bibinfo{journal}{Phys. Rev. Lett.} \textbf{\bibinfo{volume}{96}},
  \bibinfo{pages}{190401} (\bibinfo{year}{2006}).

\bibitem[{\citenamefont{DeMille}(2002)}]{DeMille02}
\bibinfo{author}{\bibfnamefont{D.}~\bibnamefont{DeMille}},
  \bibinfo{journal}{Phys. Rev. Lett.} \textbf{\bibinfo{volume}{88}},
  \bibinfo{pages}{067901} (\bibinfo{year}{2002}).

\bibitem[{\citenamefont{Balakrishnan and Dalgarno}(2001)}]{Balakrishnan01}
\bibinfo{author}{\bibfnamefont{N.}~\bibnamefont{Balakrishnan \textit{et al.}}},
  \bibinfo{journal}{Chem. Phys. Lett.} \textbf{\bibinfo{volume}{341}},
  \bibinfo{pages}{652} (\bibinfo{year}{2001}).

\bibitem[{\citenamefont{Krems}(2008)}]{Krems08}
\bibinfo{author}{\bibfnamefont{R.~V.} \bibnamefont{Krems}},
  \bibinfo{journal}{Phys. Chem. Chem. Phys.} \textbf{\bibinfo{volume}{10}},
  \bibinfo{pages}{4079} (\bibinfo{year}{2008}).

\bibitem[{\citenamefont{Sage et~al.}(2005)\citenamefont{Sage, Sainis, Bergeman,
  and DeMille}}]{Sage05}
\bibinfo{author}{\bibfnamefont{J.~M.} \bibnamefont{Sage \textit{et al.}}},
  \bibinfo{journal}{Phys. Rev. Lett.} \textbf{\bibinfo{volume}{94}},
  \bibinfo{pages}{203001} (\bibinfo{year}{2005}).

\bibitem[{\citenamefont{Ni et~al.}(2008)\citenamefont{Ni, Ospelkaus,
  de~Miranda, Pe'er, Neyenhuis, Zirbel, Kotochigova, Julienne, Jin, and
  Ye}}]{Ni08}
\bibinfo{author}{\bibfnamefont{K.-K.} \bibnamefont{Ni \textit{et al.}}},
  \bibinfo{journal}{Science} \textbf{\bibinfo{volume}{322}},
  \bibinfo{pages}{231} (\bibinfo{year}{2008}).

\bibitem[{\citenamefont{Andr\'{e} et~al.}(2006)\citenamefont{Andr\'{e},
  DeMille, Doyle, Lukin, Maxwell, Rabl, Schoelkopf, and Zoller}}]{Andre06}
\bibinfo{author}{\bibfnamefont{A.}~\bibnamefont{Andr\'{e} \textit{et al.}}},
\bibinfo{journal}{Nature Phys.} \textbf{\bibinfo{volume}{2}},
  \bibinfo{pages}{636} (\bibinfo{year}{2006}).

\bibitem[{\citenamefont{Weinstein et~al.}(1998)\citenamefont{Weinstein,
  deCarvalho, Guillet, Friedrich, and Doyle}}]{Weinstein98}
\bibinfo{author}{\bibfnamefont{J.~D.} \bibnamefont{Weinstein \textit{et al.}}},
  \bibinfo{journal}{Nature} \textbf{\bibinfo{volume}{395}},
  \bibinfo{pages}{148} (\bibinfo{year}{1998}).

\bibitem[{\citenamefont{Bethlem et~al.}(2000)\citenamefont{Bethlem, Berden,
  Crompvoets, Jongma, van Roij, and Meijer}}]{Bethlem00}
\bibinfo{author}{\bibfnamefont{H.~L.} \bibnamefont{Bethlem \textit{et al.}}},
  \bibinfo{journal}{Nature (London)} \textbf{\bibinfo{volume}{406}},
  \bibinfo{pages}{491} (\bibinfo{year}{2000}).

\bibitem[{\citenamefont{Rosa}(2004)}]{DiRosa2004}
\bibinfo{author}{\bibfnamefont{M.~D.~D.} \bibnamefont{Rosa}},
  \bibinfo{journal}{Eur. Phys. J. D} \textbf{\bibinfo{volume}{31}},
  \bibinfo{pages}{395} (\bibinfo{year}{2004}).

\bibitem[{\citenamefont{Stuhl et~al.}(2008)\citenamefont{Stuhl, Sawyer, Wang,
  and Ye}}]{Stuhl08}
\bibinfo{author}{\bibfnamefont{B.~K.} \bibnamefont{Stuhl \textit{et al.}}},
  \bibinfo{journal}{Phys. Rev. Lett.} \textbf{\bibinfo{volume}{101}}, \bibinfo{pages}{243002}
  (\bibinfo{year}{2008}).

\bibitem[{\citenamefont{Berkeland and Boshier}(2002)}]{Berkeland2002}
\bibinfo{author}{\bibfnamefont{D.~J.} \bibnamefont{Berkeland \textit{et al.}}}, \bibinfo{journal}{Phys. Rev. A}
  \textbf{\bibinfo{volume}{65}}, \bibinfo{pages}{033413}
  (\bibinfo{year}{2002}).

\bibitem[{\citenamefont{Huber and Herzberg}(1979)}]{Herzberg79}
\bibinfo{author}{\bibfnamefont{K.~P.} \bibnamefont{Huber}} \bibnamefont{and}
  \bibinfo{author}{\bibfnamefont{G.}~\bibnamefont{Herzberg}},
  \emph{\bibinfo{title}{Constants of Diatomic Molecules}}
  (\bibinfo{publisher}{Van Nostrand Reinhold}, \bibinfo{year}{1979}).

\bibitem[{\citenamefont{Dagdigian et~al.}(1974)\citenamefont{Dagdigian, Cruse,
  and Zare}}]{Dagdigian1974}
\bibinfo{author}{\bibfnamefont{P.~J.} \bibnamefont{Dagdigian \textit{et al.}}},
  \bibinfo{journal}{J. Chem. Phys.} \textbf{\bibinfo{volume}{60}},
  \bibinfo{pages}{2330} (\bibinfo{year}{1974}).

\bibitem[{\citenamefont{Allouche et~al.}(1993)\citenamefont{Allouche, Wannous,
  and Aub\'{e}rt-Frecon}}]{Allouche93}
\bibinfo{author}{\bibfnamefont{A.~R.} \bibnamefont{Allouche \textit{et al.}}},
  \bibinfo{journal}{Chem. Phys.} \textbf{\bibinfo{volume}{170}},
  \bibinfo{pages}{11} (\bibinfo{year}{1993}).

\bibitem[{\citenamefont{Ernst et~al.}(1985)\citenamefont{Ernst, K\"{a}ndler,
  Kindt, and T\"{o}rring}}]{Ernst1985}
\bibinfo{author}{\bibfnamefont{W.~E.} \bibnamefont{Ernst \textit{et al.}}},
  \bibinfo{journal}{Chem. Phys. Lett.} \textbf{\bibinfo{volume}{113}},
  \bibinfo{pages}{351} (\bibinfo{year}{1985}).

\bibitem[{\citenamefont{Steimle et~al.}(1993)\citenamefont{Steimle, Fletcher,
  and Scurlock}}]{Steimle93}
\bibinfo{author}{\bibfnamefont{T.~C.} \bibnamefont{Steimle \textit{et al.}}},
   \bibinfo{journal}{J. Mol. Spectrosc.}
  \textbf{\bibinfo{volume}{158}}, \bibinfo{pages}{487 } (\bibinfo{year}{1993}).

\bibitem[{\citenamefont{Steimle}(2008)}]{Steimle2008}
\bibinfo{author}{\bibfnamefont{T.}~\bibnamefont{Steimle}},
  \bibinfo{howpublished}{private communication} (\bibinfo{year}{2008}).

\bibitem[{\citenamefont{Brown and Carrington}(2003)}]{Brown03}
\bibinfo{author}{\bibfnamefont{J.~M.} \bibnamefont{Brown}} \bibnamefont{and}
  \bibinfo{author}{\bibfnamefont{A.}~\bibnamefont{Carrington}},
  \emph{\bibinfo{title}{Rotational spectroscopy of diatomic molecules}}
  (\bibinfo{publisher}{Cambridge Univ. Press}, \bibinfo{year}{2003}).

\bibitem[{\citenamefont{Childs et~al.}(1981)\citenamefont{Childs, Goodman, and
  Renhorn}}]{Childs81}
\bibinfo{author}{\bibfnamefont{W.~J.} \bibnamefont{Childs \textit{et al.} }},
  \bibinfo{journal}{J. Mol. Spectrosc.} \textbf{\bibinfo{volume}{87}},
  \bibinfo{pages}{522 533} (\bibinfo{year}{1981}).

\bibitem[{\citenamefont{K\"{a}ndler et~al.}(1989)\citenamefont{K\"{a}ndler,
  Martell, and Ernst}}]{Kandler89}
\bibinfo{author}{\bibfnamefont{J.}~\bibnamefont{K\"{a}ndler \textit{et al.}}},
  \bibinfo{journal}{Chem. Phys. Lett.} \textbf{\bibinfo{volume}{155}},
  \bibinfo{pages}{470 } (\bibinfo{year}{1989}).

\bibitem[{\citenamefont{Kloter et~al.}(2008)\citenamefont{Kloter, Weber,
  Haubrich, Meschede, and Metcalf}}]{Kloter2008}
\bibinfo{author}{\bibfnamefont{B.}~\bibnamefont{Kloter \textit{et al.} }},
  \bibinfo{journal}{Phys. Rev. A} \textbf{\bibinfo{volume}{77}},
  \bibinfo{eid}{033402} (\bibinfo{year}{2008}).

\bibitem[{\citenamefont{Maxwell et~al.}(2005)\citenamefont{Maxwell, Brahms,
  deCarvalho, Glenn, Helton, Nguyen, Patterson, Petricka, DeMille, and
  Doyle}}]{Maxwell05}
\bibinfo{author}{\bibfnamefont{S.~E.} \bibnamefont{Maxwell \textit{et al.}}},
  \bibinfo{journal}{Phys. Rev. Lett.} \textbf{\bibinfo{volume}{95}},
  \bibinfo{pages}{173201} (\bibinfo{year}{2005}).

\bibitem[{\citenamefont{Patterson and Doyle}(2007)}]{Patterson07}
\bibinfo{author}{\bibfnamefont{D.}~\bibnamefont{Patterson}} \bibnamefont{and}
  \bibinfo{author}{\bibfnamefont{J.~M.} \bibnamefont{Doyle}},
  \bibinfo{journal}{J. Chem. Phys.} \textbf{\bibinfo{volume}{126}},
  \bibinfo{pages}{154307} (\bibinfo{year}{2007}).

\bibitem[{\citenamefont{{Tobin} et~al.}(1987)\citenamefont{{Tobin}, {Sedgley},
  {Batzer}, and {Call}}}]{Tobin1987}
\bibinfo{author}{\bibfnamefont{A.~G.} \bibnamefont{{Tobin \textit{et al.}}}},
   \bibinfo{journal}{J. Vac. Sci. Tech.}
  \textbf{\bibinfo{volume}{5}}, \bibinfo{pages}{101} (\bibinfo{year}{1987}).

\end{thebibliography}

\end{document}